# A thermodynamic cycle more efficient than an infinite set of Carnot engines operating between the same temperature levels

José Iraides Belandria

Escuela de Ingeniería Química, Universidad de Los Andes, Mérida, Venezuela

joseiraides@ ula.ve

**Abstract**

This communication describes a theoretical thermodynamic cycle more efficient than an infinite set of Carnot engines operating between the same temperature levels. This result is unexpected from the standpoint of classical thermodynamics.

**1. Introduction**

Classical thermodynamics indicates that the most efficient thermodynamic cycle operating between two heat reservoirs is the Carnot engine [1] , and a basic theorem expresses that any reversible cycle working between two constant temperature levels should have the same efficiency as a Carnot cycle [2]. A corollary of this postulate expresses that the efficiency of any irreversible cycle is lower than that of a reversible engine operating between the same temperature levels. Also, any reversible cycle can be expressed as the limit of an infinite number of Carnot cycles [3].

To evaluate the capacity of any given thermodynamic cycle we may compare it with the efficiency of a Carnot engine working among the same temperature levels, because this engine is the most efficient of all thermodynamics cycles running between two heat reservoirs [1]. If the temperature levels are not constant, the comparison should be done with the efficiency of an infinite set of Carnot engines operating according to the temperature profile of the heat reservoirs of the cycle under study [4]. The infinite set of Carnot cycles represents an equivalent Carnot engine whose efficiency is equal to the average efficiency of the infinite set of Carnot cycles. Since the efficiency of the equivalent Carnot cycle depends only on the temperature levels of the heat reservoirs, then, once the temperature profile is defined, the average efficiency of the infinite set of Carnot cycles representing the equivalent Carnot cycle, is also fixed, whatever the choice of heat input distribution we use [5]. In other words all Carnot engines working between the same temperature levels are equivalent and have the same efficiency [2], independent of the heat distribution.

Also, conventional thermodynamics suggest that the efficiency of any cycle with not constant temperature levels should be equal or less than the efficiency of an infinite set of Carnot engines operating among the same not isothermal heat reservoirs.

In relation to this topic, the objective of this work is to show the existence of a thermodynamic cycle more efficient than an infinite set of Carnot cycles operating between the same temperature levels.

**2. The proposed thermodynamic cycle**

The cycle proposed in this work consists of three reversible stages. The first step is an isothermal expansion, the second step is a cooling process at constant volume, and the third step is an adiabatic compression.

To estimate the efficiency of this cycle let us assume that the working substance consists of 1 mol of a monatomic ideal gas enclosed in a cylinder equipped with a piston without friction. Initially, the ideal gas is at 800 K and 4 bars. Then, from this initial condition the gas expands isothermally from 4 to 1 bar. Next, the ideal gas is cooled at constant volume until it reaches an adiabatic path that returns the working substance to its initial state. To simplify calculations the ideal gas is considered as a closed system and the cylinder walls and the piston are assumed of negligible heat capacity. Also kinetic and potential energy changes are neglected.

We can find, using standard thermodynamic calculations, that during the reversible isothermal expansion the gas absorbs 9220.52 J of heat from the high temperature reservoir which is at a constant temperature of 800 K, and produces 9220.52 J of work. For this to be a reversible heat transfer the temperature of the working fluid must be infinitesimally lower than the temperature of the hot reservoir. During the reversible constant volume step, the ideal gas is cooled from 800 K to 317.48 K and rejects 6017.51 J of heat to an infinite set of cold heat reservoirs whose temperatures varies from 800 K to 317.48 K. During this process the temperature of the working substance is infinitesimally higher than the temperature of the cold reservoirs. Next, the ideal gas requires 6017.51 J of reversible compression work to return adiabatically to its initial state. From these calculations, the energy supplied to the proposed cycle is 9220.52 J and the net work is 3203.01 J. Then, the thermal efficiency of the cycle is 0.35. The following equation expresses the thermal efficiency η in terms of the absolute temperatures Th and Tc.

$$\eta = 1 - [ ( Th - Tc ) / Ln ( Th / Tc ) ] / Th \qquad (1)$$

where Th is the temperature of the hot reservoir and Tc is the lowest temperature of the cold reservoir.

Energy and entropy balances suggest that energy is conserved and that the total entropy change of the given cycle is 0 cal / K.

### 3. The infinite set of Carnot cycles

Now, we may compare the previous thermodynamic cycle with an infinite set of Carnot engines operating between the same heat reservoirs. As we have observed, the hot reservoir is at a constant temperature of 800 K, but the temperature of the cold reservoir is not constant and varies from 800 K to 317.48 K. Therefore, it is not possible to use only one Carnot engine to operate among these heat reservoirs; instead, we need an infinite set of parallel Carnot cycles working between the above heat reservoirs.

Then, let us remove the proposed cycle from the specified heat reservoirs and substitute it for an infinite set of parallel Carnot cycles that absorb heat from the high temperature reservoir which is at a constant temperature of 800 K, and expel heat to an infinite set of low temperature reservoirs ranging uniformly from 800 K to 317.48. As an example, the first Carnot cycle of the set receives a certain amount of heat from the hot reservoir at 800 K, produces work and discards some heat to the cold reservoir at 317.47 K. The second Carnot cycle of the set takes a portion of heat from the hot reservoir at 800 K, does work and rejects some heat to the cold reservoir at 317.48 K, for instance. The third Carnot cycles takes a part of heat from the hot reservoir at 800 K, performs work and releases heat to a cold reservoir at 317.50 K. And so on, we assemble an infinite set of parallel Carnot engines to cover the whole range of the cold temperature reservoirs ranging uniformly from 800 K to 317.48 K. If the number of parallel Carnot cycles tends to infinite, the average efficiency of this infinite set of Carnot engines will match with the efficiency of an equivalent Carnot engine working between the specified heat reservoirs.

The average efficiency of this infinite set of parallel Carnot cycles, $\eta_{av}$, is equal to the net work of the infinite set of Carnot cycles Wnet, divided by the total heat absorbed by the set of Carnot engines Qh. Thus,

$$\eta_{av} = Wnet / Qh \qquad (2)$$

And

$$Wnet = W_1 + W_2 + W_3 + \ldots\ldots + Wn \qquad (3)$$

$$Qh = Qh_1 + Qh_2 + Qh_3 + \ldots + Qhn \qquad (4)$$

Here, $W_1$, $W_2$, $W_3$, $W_n$ represent the net work of each Carnot cycle of the set, and $Qh_1$, $Qh_2$, $Qh_3$, $Qh_n$ are the heat absorbed from the high temperature reservoir by each Carnot engine. The sub index n symbolizes the $n^{th}$ Carnot cycle and varies from 1 to infinite.

Now, the efficiency of the $n^{th}$ Carnot engine of the set $\eta_n$ is

$$\eta_n = Wn / Qhn = (1 - Tcn / Th) \quad (5)$$

where Th is the temperature of the hot reservoir of each Carnot cycle which is the same for all of the cycles of the set and equal to 800K, and Tcn is the temperature of the cold reservoir of the $n^{th}$ Carnot cycle. Tcn varies from 317.48 K to 800 K.

Then, the work of the $n^{th}$ Carnot cycle is

$$Wn = (1 - Tcn / Th) / Qhn \quad (6)$$

Subsequently, the net work of the infinite set of Carnot cycles is

$$Wnet = (1 - Tc_1 / Th) Qh_1 + (1 - Tc_2 / Th) Qh_2 + \ldots + (1 - Tcn / Th) Qhn \quad (7)$$

By combining Eqs (4) and (7)

$$Wnet = Qh - [Tc_1 Qh_1 + Tc_2 Qh_2 + \ldots + Tcn Qhn] / Th \quad (8)$$

Or, in a compact notation

$$Wnet = Qh - \left( \lim_{n \to \infty} \sum_{i=1}^{n} Tci\, Qhi \right) / Th \quad (9)$$

To get the limit of Eq. (9) we may employ equal or different values of Qhi. In any case, the result will be the same if each Qhi tends to zero [6]. This fact support the argument, previously discussed, that suggest that the average efficiency of the infinite set of Carnot cycles does not depends on the amount of heat received by each Carnot cycle. Then, for simplicity, we may use a uniform value for each Qhi. Thus

$$Qh_1 = Qh_2 = Qhi = Qh / n \quad (10)$$

Since n tends to infinite, Qhi tends to zero, and the limit of Eq.(9) converges to a unique value.

By substituting Eq. (10) into Eq. (9) it is found that

$$Wnet = Qh \left[ 1 - \left( \lim_{n \to \infty} \sum_{i=1}^{n} Tci / n \right) / Th \right] \quad (11)$$

From the general concept of integration [6], the above limit coincides with the integral of the following expression

$$\lim_{n \to \infty} \sum_{i=1}^{n} Tci / n = [1 / (Tcb - Tca)] \int_{Tca}^{Tcb} Tc\, dTc = (Tca + Tcb) / 2 \quad (12)$$

Here, Tca is the lowest temperature of the cold reservoir equal to 317.48 K and Tcb is the highest temperature of the cold reservoir equal to 800 K .

By combining Eqs (11) and (12), it is found that

$$\text{Wnet} = \text{Qh} \ [\ 1 - (\text{Tca} + \text{Tcb})\ /\ 2\ \text{Th}\ ] \tag{13}$$

And, by introducing Eq. (13) into Eq. (2), we obtain

$$\eta_{av} = 1 - [(\text{Tcb} + \text{Tca})\ /\ 2\ \text{Th}\ ] \tag{14}$$

Then, the average efficiency $\eta_{av}$ of the infinite set of Carnot cycles is 0.30.

By evaluating Eq.(14) with an alternative procedure [7] by using the limit of a series, the same result is obtained for the average efficiency. Also, a method proposed in example (13-1) of reference [4] by applying a differential expression to determine the net work of a set of Carnot engines gives an equal result.

## 4. Conclusion

We observe that the efficiency of the cycle described in this work is greater than the efficiency of the infinite set of Carnot cycles operating between the same temperature levels. This result is unforeseen from the perspective of classical thermodynamics which suggests that the efficiency of any thermodynamic cycle with no isothermal temperature levels should be equal or less than the efficiency of an infinite set of Carnot engines working with the same heat reservoirs

.

## 5. References

[1] Zemansky M.W. and Van Ness H.C., Basic Engineering Thermodynamics, McGraw Hill Inc., New York, 1966.

[2] Perrot P., A to z of Thermodynamics, Oxford University Press Inc., New York, 1998.

[3] Dickerson R., Molecular Thermodynamics, W.A. Benjamin Inc., Menlo Park, California,1969.

[4] Smith J.M. and Van Ness H.C., Introduction to Chemical Engineering Thermodynamics, 3 th edition, McGraw Hill Inc., New York, 1975

[5] Work of an equivalent Carnot cycle, Advanced Physics Forums http:// www. advancedphysics.org/viewthread.php?tid=1881.

[6] Sokowsky E.W., Calculus with Analytic Geometry, 4th edition, PWS Publishers, USA, 1988.

[7] Eficiency of an infinite series of Carnot cycles, Physics forums http:// www.physicsforums.com/.